\documentclass[runningheads]{llncs}
\usepackage[T1]{fontenc}
\usepackage{graphicx}
\usepackage{amsmath}
\usepackage{amssymb}
\usepackage[dvipsnames]{xcolor}
\usepackage{soul}
\usepackage{booktabs}
\usepackage{dsfont}
\usepackage{fontawesome5}
\usepackage{paralist}
\usepackage{multirow}
\usepackage{colortbl}

\definecolor{lg}{HTML}{e7e7e7}

\newcommand{\ra}{$\rightarrow$}

\graphicspath{{./figs/}}

\begin{document}

\title{Self-supervised learning via inter-modal reconstruction and feature projection networks for label-efficient 3D-to-2D segmentation}

\titlerunning{Self-supervised learning method for label-efficient 3D-to-2D segmentation}
%

\author{%
    José Morano\inst{1}\textsuperscript{(\faEnvelope[regular])}
    \and
    Guilherme Aresta\inst{1}
    \and
    Dmitrii Lachinov\inst{1}
    \and
    Julia Mai\inst{2}
    \and
    Ursula Schmidt-Erfurth\inst{2}
    \and
    Hrvoje Bogunović\inst{1,2}
}
%
%
\authorrunning{J. Morano et al.}
%
\institute{
Christian Doppler Laboratory for Artificial Intelligence in Retina, Department of Ophthalmology and Optometry, Medical University of Vienna, Austria
\and
Lab for Ophthalmic Image Analysis, Department of Ophthalmology and Optometry, Medical University of Vienna, Austria\\
\texttt{\{jose.moranosanchez,hrvoje.bogunovic\}@meduniwien.ac.at}
}

\maketitle              

\begin{abstract}
Deep learning has become a valuable tool for the automation of certain medical image segmentation tasks, significantly relieving the workload of medical specialists.
Some of these tasks require segmentation to be performed on a subset of the input dimensions, the most common case being 3D\ra2D.
However, the performance of existing methods is strongly conditioned by the amount of labeled data available, as there is currently no data efficient method, e.g. transfer learning, that has been validated on these tasks.
In this work, we propose a novel convolutional neural network (CNN) and self-supervised learning (SSL) method for label-efficient 3D\ra2D segmentation.
The CNN is composed of a 3D encoder and a 2D decoder connected by novel 3D\ra2D blocks.
The SSL method consists of reconstructing image pairs of modalities with different dimensionality.
The approach has been validated in two tasks with clinical relevance: the en-face segmentation of geographic atrophy and reticular pseudodrusen in optical coherence tomography.
Results on different datasets demonstrate that the proposed CNN significantly improves the state of the art in scenarios with limited labeled data by up to 8\% in Dice score.
Moreover, the proposed SSL method allows further improvement of this performance by up to 23\%, and we show that the SSL 
is beneficial regardless of the network architecture.
Our code is available at \texttt{\url{https://github.com/j-morano/multimodal-ssl-fpn}}.

\end{abstract}


\setcounter{footnote}{0}

\section{Introduction}

Deep learning can significantly reduce the workload of medical specialists during image segmentation tasks, which are essential for patient diagnosis and follow-up management~\cite{Kavur_MIA_2021,Menze_BRATS_TMI_2015,Orlando_MIA_2020}.
For most tasks, segmentation masks have the same dimensionality as the input.
However, there are some tasks for which segmentation has to be performed in a subset of the dimensions of the data, e.g. 3D$\rightarrow$2D~\cite{Liefers_MIDL_2019,Sun_TMI_2013}.
This occurs, for example, for the segmentation of geographic atrophy (GA) in optical coherence tomography (OCT), where the segmentation is performed on the OCT projection.
In recent years, several methods have been proposed for this type of tasks~\cite{Lachinov_MICCAI_2021,Li_IPN_TMI_2020,Li_IPNv2_arXiv_2020,Liefers_MIDL_2019}.
Li \textit{et al.}~\cite{Li_IPN_TMI_2020} proposed an image projection network (IPN) that reduces the features to the target dimensionality using unidirectional pooling layers in the encoder.
However, IPN follows a patch-based approach with fixed patch size, which prevents its direct application to full 3D volumes of varying size.
Also, it does not have skip connections, which have proven to be highly useful for accurate segmentation.
Later, Lachinov \textit{et al.}~\cite{Lachinov_MICCAI_2021} proposed a U-Net-like convolutional neural network (CNN) for 3D$\rightarrow$2D segmentation that overcomes the limitations of IPN, which were also later overcome by the second version of IPN (IPNv2)~\cite{Li_IPNv2_arXiv_2020}.
However, they still require a large amount of labeled data to provide adequate performance.
In addition, there are works that explore the use of CNNs for 3D\ra2D regression, where 
Seeböck \textit{et al.}~\cite{Seebok_OR_2022} proposed ReSensNet, a novel CNN based on Residual 3D U-Net~\cite{Lee_Res3DUNet_arXiv_2017}, with a 3D encoder and a 2D decoder connected by 3D\ra2D blocks.
However, ReSensNet only works at concrete input resolutions, and it is applied pixel-wise.

In general, one of the issues of these and other deep learning segmentation methods is that their performance strongly depends on the amount of annotated data~\cite{Tajbakhsh_MIA_2020}, which hinders their deployment to real-world medical image analysis settings.
Transfer learning from ImageNet is the standard approach to mitigate this issue~\cite{Tajbakhsh_MIA_2020}.
However, specifically for segmentation, ImageNet pre-training has shown minimal performance gains~\cite{He_ICCV_2019,Raghu_NeurIPS_2019}, partially because it can only be performed on the encoder part of the very common encoder-decoder architectures.

A possible alternative is to pre-train the models using a self-supervised learning (SSL) paradigm~\cite{Kalapos_ECCVW_2022,Brempong_Denoising_CVPR_2022,Grill_NIPS_2020,Hervella_MICCAI_2018,Hervella_ASOC_2020,Morano_ECAI_2020,Ross_IJCARS_2018}.
However, only some of these approaches have the potential to be applied for 3D\ra2D segmentation, as many of them, such as image denoising~\cite{Brempong_Denoising_CVPR_2022}, require input and output images to have the same dimensionality.
Among the suitable approaches, multi-modal reconstruction pre-training (MMRP) shows great potential in multi-modal scenarios~\cite{Hervella_ASOC_2020}.
In this approach, models are trained to reconstruct pairs of images from different modalities, learning relevant patterns in the data without requiring manual annotations.
MMRP, however, has only been proven useful for localizing non-pathological structures on 2D color fundus photography, using fluorescein angiography as the modality to reconstruct.
Moreover, image pairs of these modalities have to be registered using a separate method.

\paragraph{Contributions.} 
In this work, we propose a novel approach for label-efficient 3D$\rightarrow$2D segmentation.
In particular, our contributions are as follows:
\begin{inparaenum}[(1)]
\item As an alternative to state-of-the-art network architectures, we propose a 3D$\rightarrow$2D segmentation CNN based on ReSensNet~\cite{Seebok_OR_2022} that has a 3D encoder and a 2D decoder connected by novel 3D\ra2D projective blocks.
\item We propose a novel SSL strategy for 3D\ra2D models based on the reconstruction of modalities of different dimensionality, and show that it significantly improves the performance of the models in the target tasks.
This is the first data efficient method proposed for 3D\ra2D models and the first work exploring 3D\ra2D reconstruction.
\item Lastly, the performed experiments deepen the understanding of the proposed SSL paradigm, by exploring different settings with different image modalities.
\end{inparaenum}
The proposed approach was validated on two clinically-relevant tasks: the en-face segmentation of GA and reticular pseudodrusen (RPD) in retinal OCT.
The results demonstrate that the proposed approach clearly outperforms the state of the art in scenarios with scarce labeled data.

\subsubsection{Clinical background.} 
\begin{figure}[tbp]
\centering
\begin{tabular}{ccccc}
\includegraphics[width=0.38\textwidth]{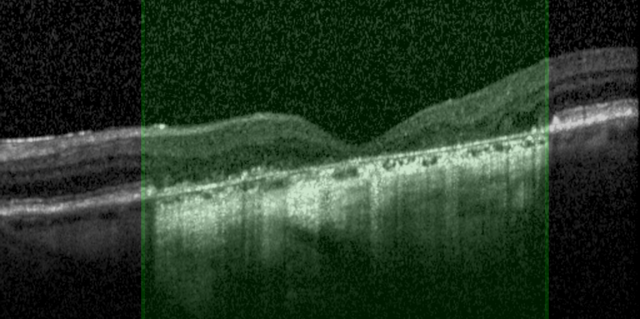}
& \includegraphics[width=0.19\textwidth]{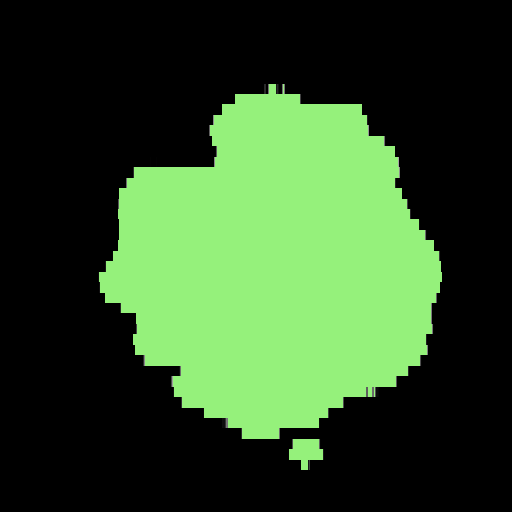}
& \includegraphics[width=0.19\textwidth]{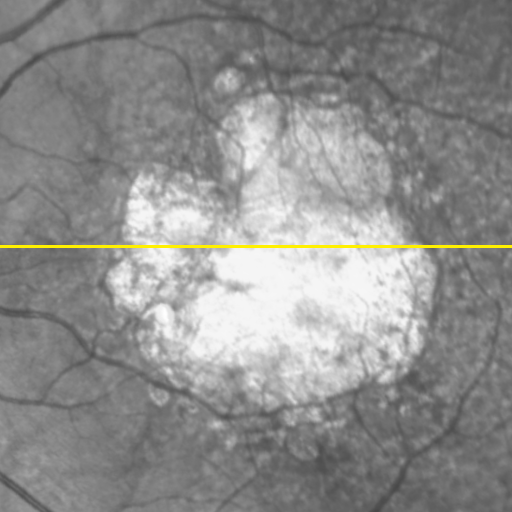}
& \includegraphics[width=0.19\textwidth]{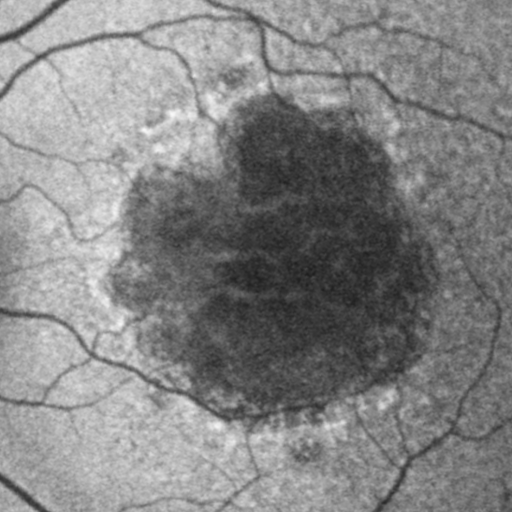} \\

\includegraphics[width=0.38\textwidth]{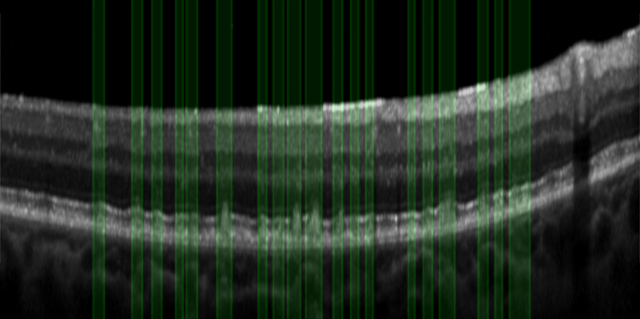}
& \includegraphics[width=0.19\textwidth]{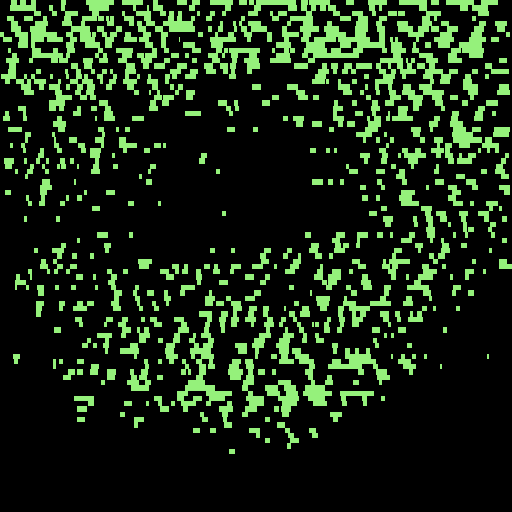}
& \includegraphics[width=0.19\textwidth]{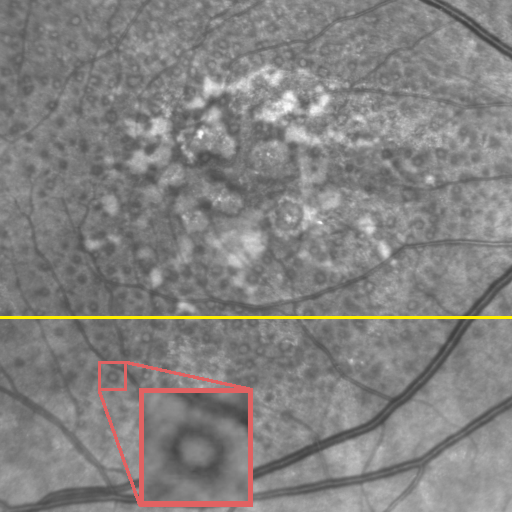}
&
\end{tabular}
\caption{%
From left to right: OCT slice (B-scan) with the corresponding ground truth annotations overlaid in green, ground truth, SLO with the location of the B-scan indicated in yellow and a zoom-in view in red, and FAF.
Top: GA.
Bottom: RPD.
}%
\label{fig:clinical_background}
\end{figure}

\emph{Geographic atrophy} (GA) is an advanced form of age-related macular degeneration (AMD) that corresponds to a progressive loss of retinal photoreceptors and leads to irreversible visual impairment.
GA is typically assessed with OCT and/or fundus autofluorescence (FAF) imaging modalities~\cite{Bui_Eye_2021,Wei_Eye_2023}.
In OCT, it is characterized by the loss of retinal pigment epithelium (RPE) tissue, accompanied by the contrast enhancement of the signal below the retina, and in FAF, by the loss of RPE autofluorescence~\cite{SchmitzValckenberg_O_2016} (see Fig.~\ref{fig:clinical_background}).
In both cases, GA lesion is delineated as a 2D en-face area.
Also, GA frequently appears brighter than the surrounding areas on scanning laser ophthalmoscopy (SLO) images due to its higher reflectance.

\emph{Reticular pseudodrusen} (RPD) are accumulations of extracellular material that commonly occur in association with AMD.
In OCT scans, these lesions are shown as granular hyperreflective deposits situated between the RPE layer and the ellipsoid zone.
SLO visualizes RPD as a reticular pattern of iso-reflective round lesions surrounded by a hyporeflective border (see Fig.~\ref{fig:clinical_background}).

\section{Methods and experimental setup}

\begin{figure}[tbph]
\centering
     \includegraphics[width=\textwidth]{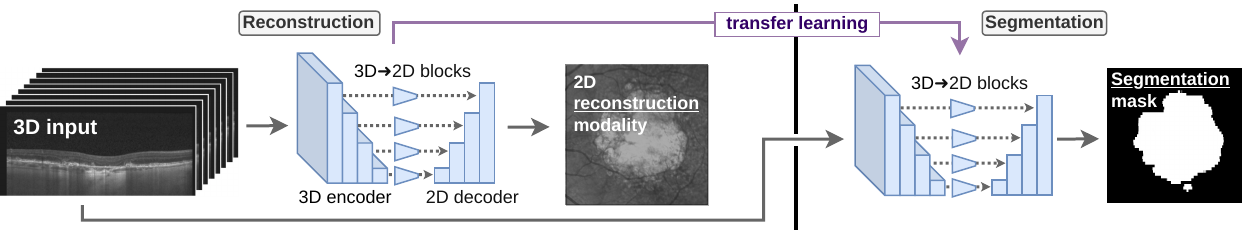}
\caption{%
Illustration of the proposed approach for 3D\ra2D segmentation.
A novel 3D\ra2D model is trained for reconstructing image pairs of modalities with different dimensionality in a SSL setting, and then fine-tuned in the target segmentation task.}
\label{fig:overall}
\end{figure}

The proposed approach, illustrated in Fig.~\ref{fig:overall}, is as follows.
Let $\textbf{x} \in \mathcal{X} \subset \mathbb{R}^n$ and $\textbf{y} \in \mathcal{Y} \subset \mathbb{R}^{n-1}$ be two images from modalities $\mathcal{X}$ and $\mathcal{Y}$, and $\textbf{z} \in \mathcal{Z} \subset \mathbb{R}^{n-1}$, their corresponding target segmentation mask.
Let images and masks from $\mathcal{Y}$ and $\mathcal{Z}$ have related anatomical features.
We optimize a reconstruction model $\textbf{y} = f_r(\textbf{x};\theta_r)$ and transfer its knowledge by initializing the weights of the segmentation model $\textbf{z} = f_s(\textbf{x};\theta_s)$ with the optimized weights of the reconstruction model $f_r$.
With this approach, modality $\mathcal{Y}$ serves as a free source of supervision, and images of this modality serve as soft segmentation targets.
Thus, models can learn relevant patterns in a self-supervised way.

In this work, we propose a new CNN for the special case of 3D\ra2D segmentation, and we evaluate the proposed SSL approach for this case.
In particular, we pre-train the new CNN to reconstruct SLO/FAF images from OCT (3D$\rightarrow$2D), and then fine-tune it for GA/RPD segmentation.
The advantage of reconstructing SLO over FAF is that several modern OCT devices allow to obtain co-registered OCT and SLO scans, providing the coordinates of each OCT slice within the SLO; thus, there is no need to use a separate registration method.

\subsubsection{Network architecture.}
\begin{figure}[tbph]
\centering
\includegraphics[width=\textwidth]{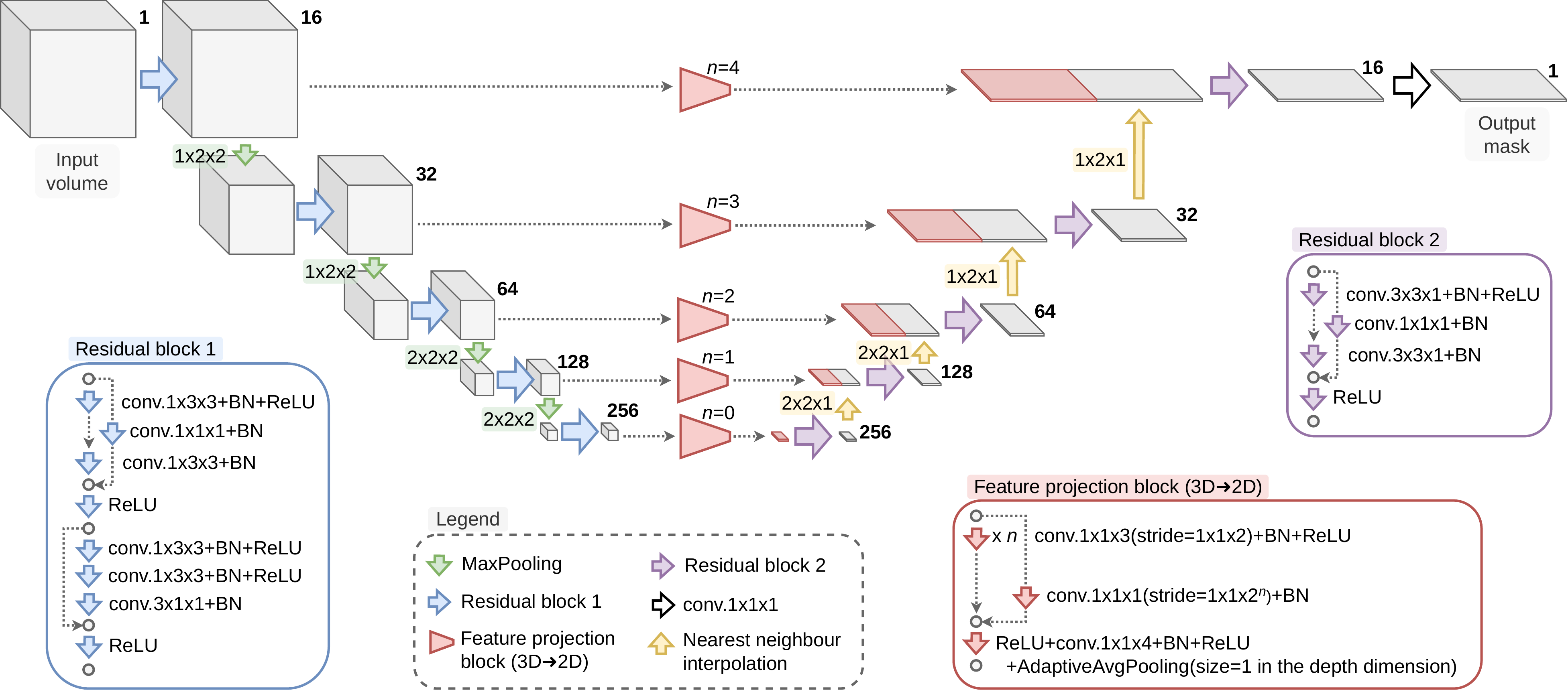}
\caption{Proposed 3D$\rightarrow$2D CNN.
Each residual encoder block has 8 3D convolutional layers, and each residual decoder block has 4 2D layers (number of feature maps also shown).
The proposed feature projection block (FPB, in red) projects 3D features to the 2D feature space.
FPBs have a variable number of $1 \times 1 \times 3$ convolutions followed by a $1 \times 1 \times 4$ convolution and a depth-wise adaptive average pooling of size 1.
}
\label{fig:architecture}
\end{figure}

The proposed network architecture (Fig.~\ref{fig:architecture})
is based on ReSensNet~\cite{Seebok_OR_2022}, and consists of a 3D encoder and a 2D decoder connected by novel 3D\ra2D feature projection blocks (FPBs).
In the original work~\cite{Seebok_OR_2022}, training and inference are performed pixel-wise using fixed-size input patches.
In contrast, we use full-size volumes of arbitrary resolution.
To this end, we propose a novel type of FPB.
In particular, all convolutions whose kernel size was equal to the expected feature size (calculated from the fixed size of the input patch)  were replaced by $1 \times 1 \times 4$ convolutions.
Then, to project 3D features at the output of each FPB to the 2D feature space, we add an adaptive average pooling of size 1 in the depth dimension at the end of each block.
With this setting, feature selection and dimension reduction are performed at different scales, and the decoder processes only 2D features in the selected dimensions.
This allows the model to learn the 3D structure of the data while being able to perform the segmentation in 2D.
In addition, to overcome memory constraints and avoid overfitting, we reduce by half the number of kernels in each convolutional block.

\subsubsection{Training losses.}
As \emph{reconstruction loss}, we use negative mean structural similarity index (NMSSIM)~\cite{Wang_SSIM_TIP_2004}.
We empirically found that this loss performs equally or better than modern perceptual losses (e.g. LPIPS~\cite{Zhang_LPIPS_CVPR_2018}) for our approach.
NMSSIM loss can be defined as
$\mathcal{L}_{\text{NMSSIM}}(\textbf{x},\textbf{y}) = -\frac{1}{HW}\sum_{h,w} \text{SSIM}(\textbf{x}_{hw}, \textbf{y}_{hw})$,
where $\textbf{x}_{hw}$ and $\textbf{y}_{hw}$ denote image patches of images $\textbf{x}$ and $\textbf{y}$ centered on the pixel with coordinates $(h,w)$, $h \in H$ and $w \in W$, and $\text{SSIM}(\textbf{x}_{hw}, \textbf{y}_{hw})$ is the SSIM map for those patches, as described in~\cite{Wang_SSIM_TIP_2004}.
As \emph{segmentation loss}, we use the direct sum of Dice loss and Binary Cross-Entropy.
These two losses are standard for binary segmentation tasks~\cite{Kavur_MIA_2021,Lachinov_MICCAI_2021,Li_IPN_TMI_2020,Li_IPNv2_arXiv_2020,Orlando_MIA_2020}.

\subsubsection{Datasets.}

Experiments were performed using three datasets (Table~\ref{tab:data_summary}).
\begin{table}[tbp]
\setlength{\tabcolsep}{6pt}
\caption{
Study data details.
Dimensions: en-face height $\times$ en-face width $\times$ depth. FAF and SLO characteristics are after cropping to the same OCT en-face region projection.
}%
\label{tab:data_summary}
\centering
\resizebox{\textwidth}{!}{%
\begin{tabular}{lccccccc}

\toprule
\textbf{Dataset}
& \textbf{Scans}
& \textbf{Patients (eyes)}
& \textbf{Modality}
& \textbf{Device}
& \textbf{Area} (mm)
& \textbf{Size} (px) \\

\midrule
GA-M
& 967
& 100 (184)
& \begin{tabular}[c]{@{}l@{}}
        OCT \\
        SLO \\
        FAF
    \end{tabular}
& \begin{tabular}[c]{@{}l@{}}
        Spectralis \\
        Spectralis \\
        Spectralis
    \end{tabular}
& \begin{tabular}[c]{@{}l@{}}
        $6.68 \times 6.68 \times 1.92$ \\
        $6.68 \times 6.68$ \\
        $6.68 \times 6.68$
    \end{tabular}
& \begin{tabular}[c]{@{}l@{}}
        $49 \times 1024 \times 496$ \\
        $1024 \times 1024$ \\
        $1024 \times 1024$
    \end{tabular} \\

\midrule
GA-S
& 270
& 149 (166)
& OCT
& Spectralis
& $6.02 \times 6.03 \times 1.92$
& $49 \times 512 \times 496$ \\

\midrule
RPD-S
& 23
& 19 (23)
& OCT
& Spectralis
& $5.73 \times 5.72 \times 1.92$
& $97 \times 1024 \times 496$ \\

\bottomrule
\end{tabular}
}
\end{table}
\emph{GA-M}
samples come from a clinical study on GA progression.
OCT and SLO images were automatically co-registered by the imaging device, while FAF images were registered with SLO using an in-house pipeline based on aligning retinal vessel segmentation.
FAF and SLO images were cropped and resized to the same area and resolution as the OCT en-face projection.
GA-M-S (35 samples) is a subset of GA-M with GA en-face masks annotated by a retinal expert on the OCT B-Scans. 
\emph{GA-S}
is composed of OCT volumes from another study with en-face GA annotations created by a retinal expert on the OCT B-scans.
This dataset is divided patient-wise into two subsets: GA-S-2, containing volumes with annotations of two different experts, and GA-S-1, of only one.
\emph{RPD-S}
is composed of OCT volumes with en-face RPD annotations created by retinal experts.

\subsubsection{Training and evaluation details.}
OCT volumes were flattened along the Bruch's membrane, rescaled depth-wise to 128 voxels, and then Z-score normalized along the cross-sectional plane.
To make FAF and GA masks more similar and thus facilitate fine-tuning, FAF images were inverted.
In all cases, models were trained for 800 epochs using SGD with a learning rate of $0.1$ and a momentum of $0.9$.
Batch size was set to 4 for reconstruction and 8 for segmentation.

All datasets were split patient-wise into training (60\%), validation (10\%) and test (30\%).
For reconstruction, models were trained on GA-M.
For GA segmentation, they were trained/fine-tuned on GA-S-1 and evaluated on GA-S-1, GA-S-2 and GA-M-S.
For RPD segmentation, RPD-S was used.
To evaluate the performance under label scarcity, we train with 5\%, 10\%, 20\% and 100\% of the data in GA-S, and 20\% and 100\%, in RPD-S.
More details about the hardware used and the carbon footprint of our method are included in the Supplement.

To reduce inference variability, we average the predictions of the top-5 checkpoints of the models in terms of Dice (validation).
Segmentations are evaluated via Dice and absolute area difference (Area diff.) of predicted and manual masks.

\section{Results and discussion}

\begin{figure}
\centering
\includegraphics[height=0.34\textwidth]{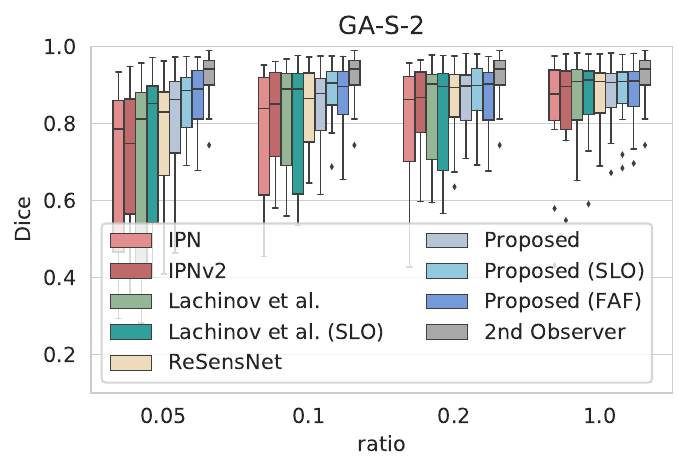}
\hfill
\includegraphics[height=0.34\textwidth]{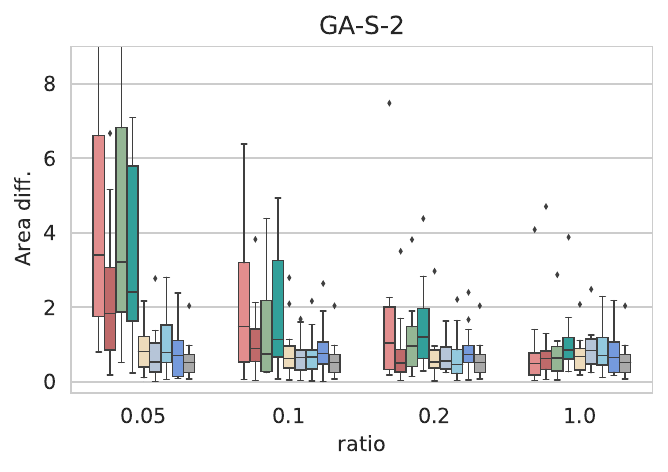}
\\
\includegraphics[height=0.34\textwidth]{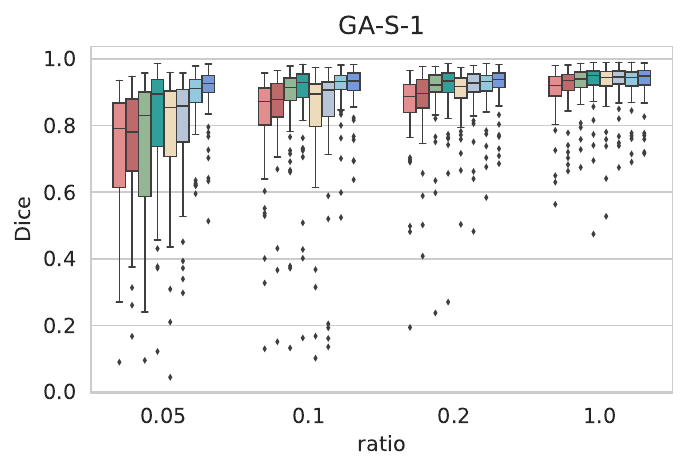}
\hfill
\includegraphics[height=0.34\textwidth]{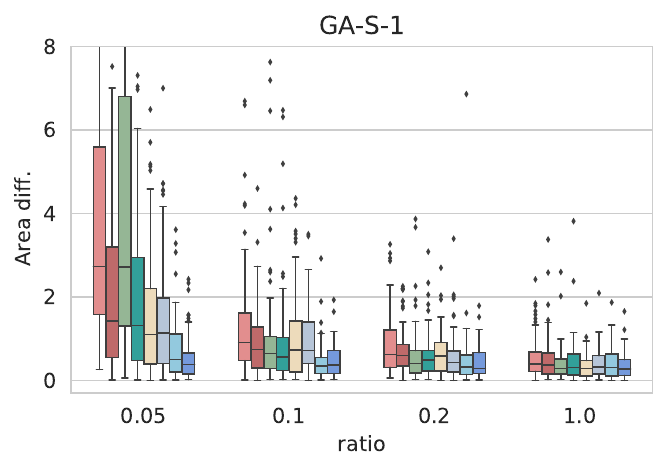}
\\
\includegraphics[height=0.34\textwidth]{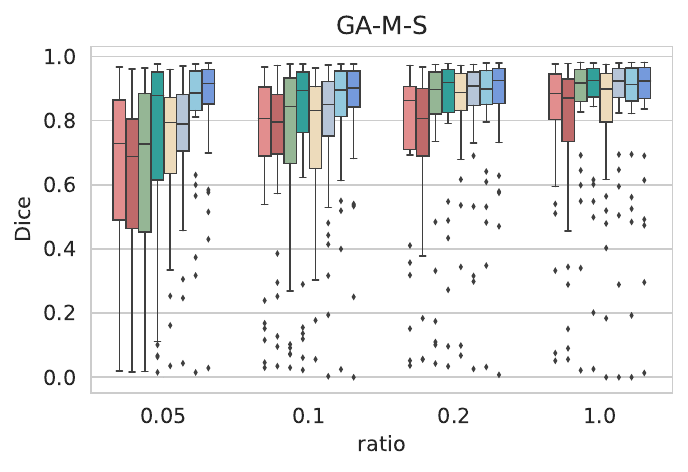}
\hfill
\includegraphics[height=0.34\textwidth]{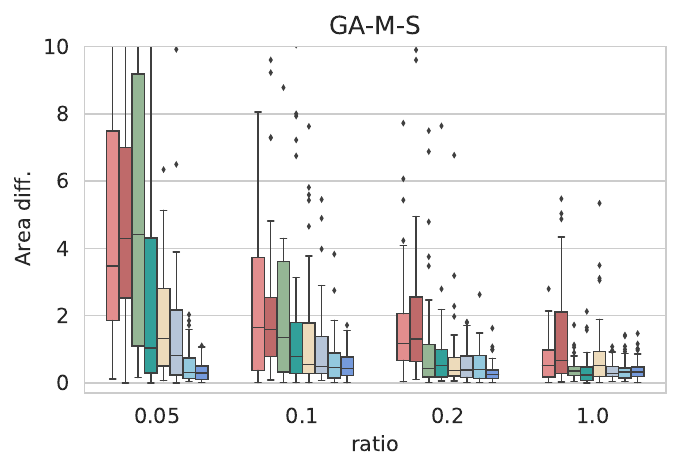}
\\
\includegraphics[height=0.31\textwidth]{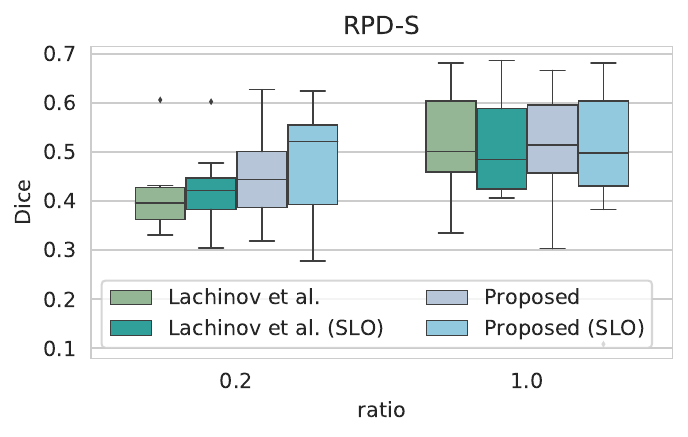}
\hfill
\includegraphics[height=0.31\textwidth]{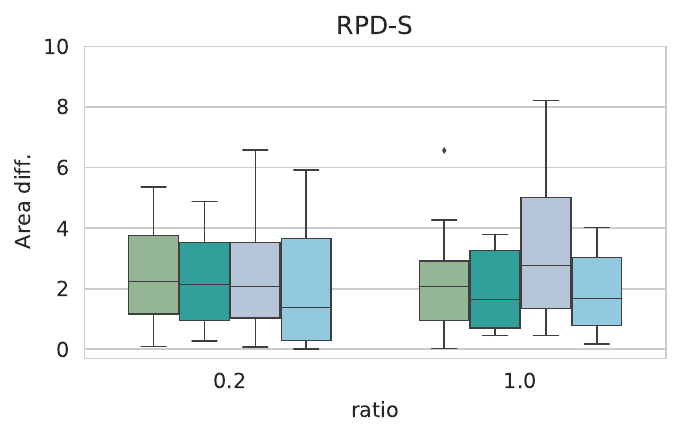}
\caption{%
Segmentation results of the models trained with different amounts of data.
The title of each plot indicates the test dataset.
If a model was pre-trained with SSL, the pre-training modality is shown in parentheses. 
A table with all means and standard deviations, as well as the results of a Wilcoxon signed rank test between our proposal and the others is included in the Supplement.
}%
\label{fig:results_models}
\end{figure}
\begin{figure}
\centering
\scriptsize
\begin{tabular}{cccccc}
\includegraphics[width=0.155\textwidth]{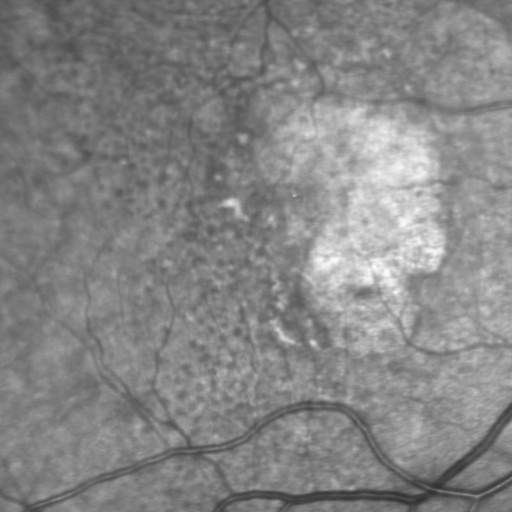}
& \includegraphics[width=0.155\textwidth]{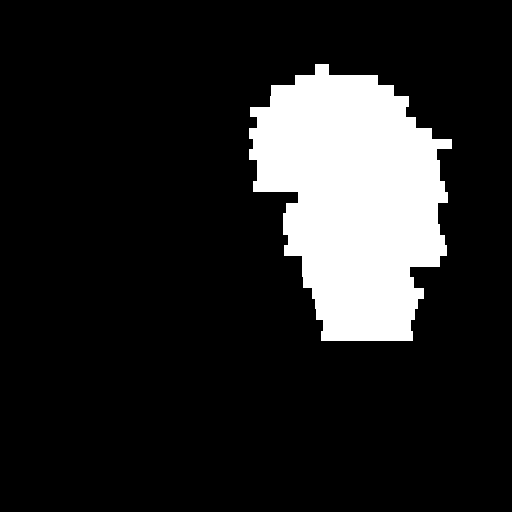}
& \includegraphics[width=0.155\textwidth]{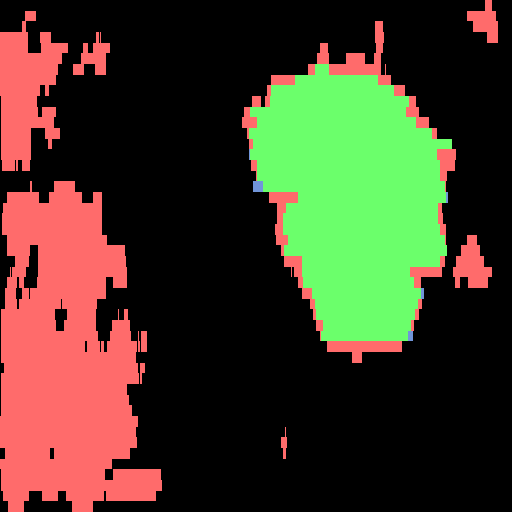}
& \includegraphics[width=0.155\textwidth]{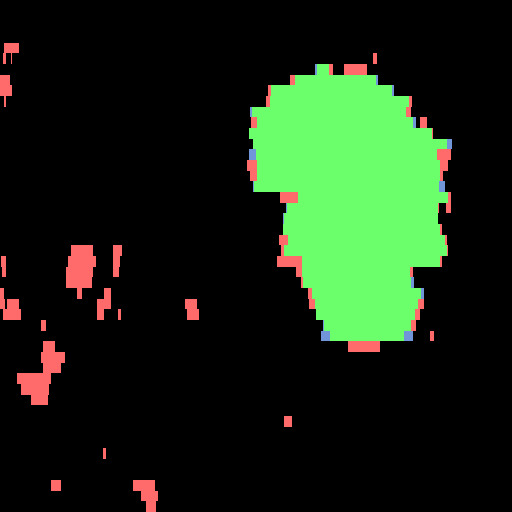}
& \includegraphics[width=0.155\textwidth]{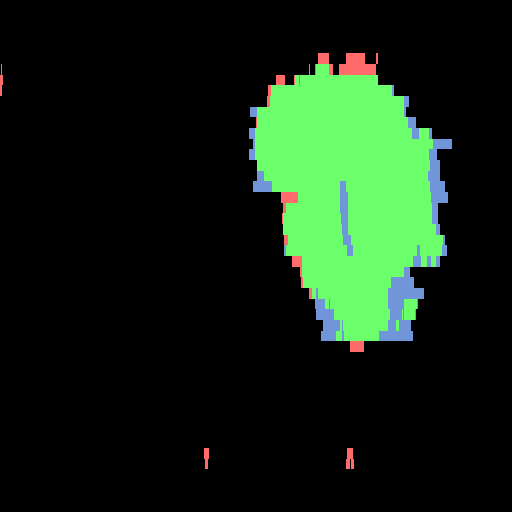}
& \includegraphics[width=0.155\textwidth]{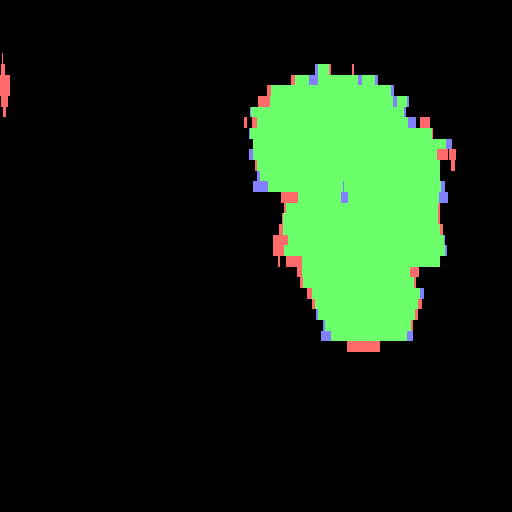} \\
\includegraphics[width=0.155\textwidth]{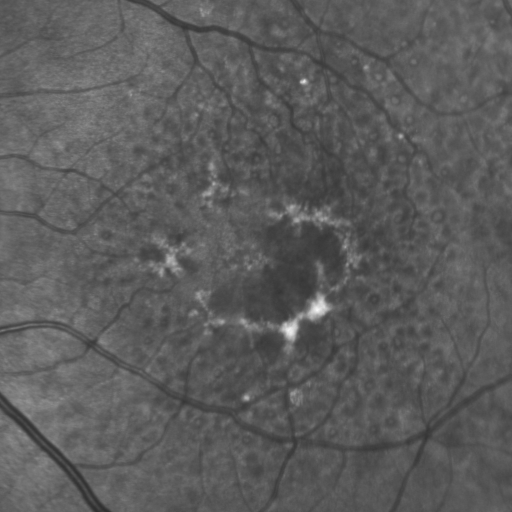}
& \includegraphics[width=0.155\textwidth]{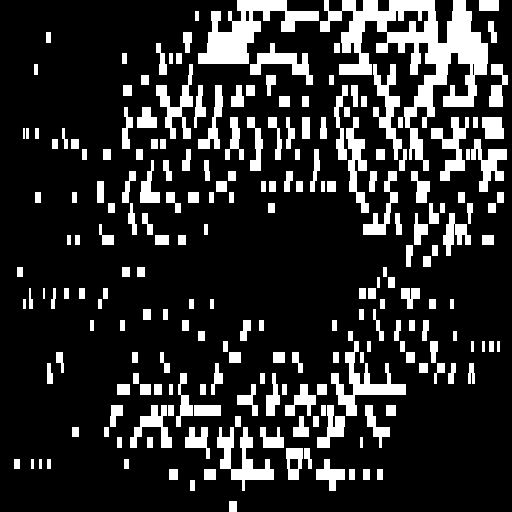}
& \includegraphics[width=0.155\textwidth]{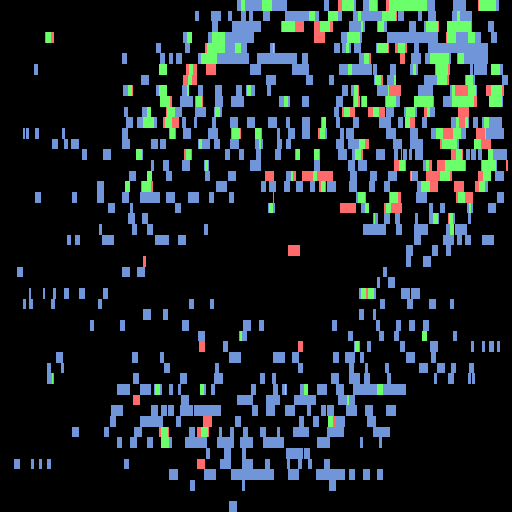}
& \includegraphics[width=0.155\textwidth]{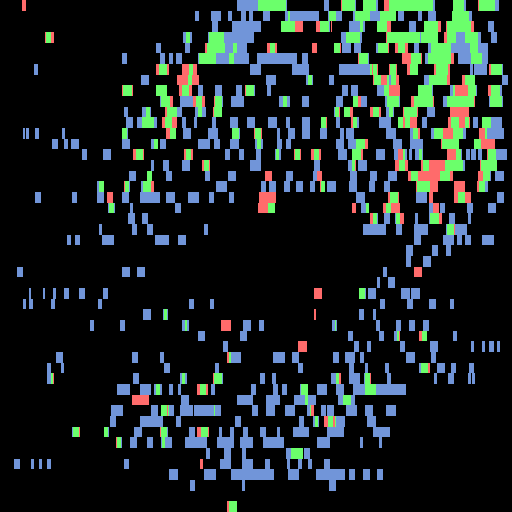}
& \includegraphics[width=0.155\textwidth]{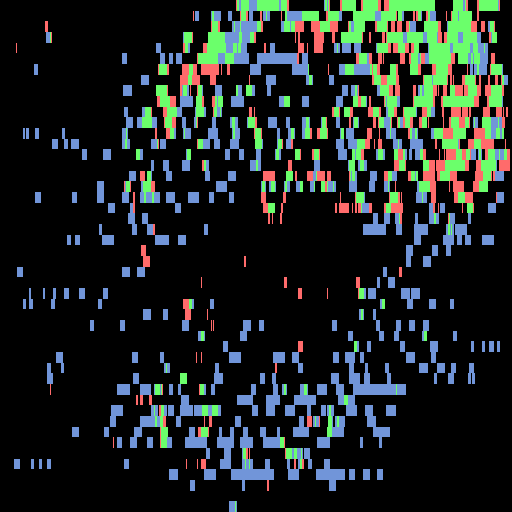}
& \includegraphics[width=0.155\textwidth]{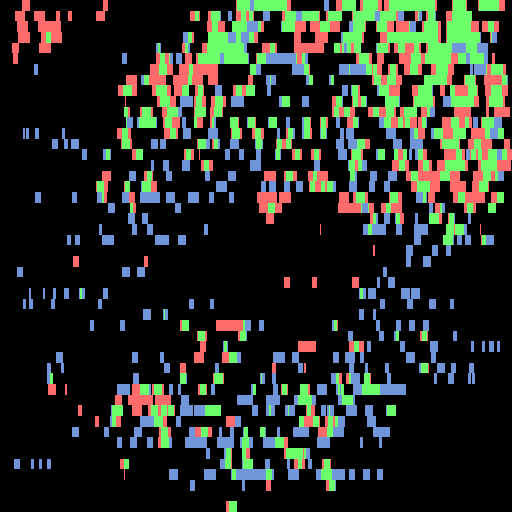} \\
SLO
& Ground
& Lachinov
& Lachinov
& Proposed
& Proposed \\

& truth
& \textit{et al}.
& \textit{et al}. (SLO)
& 
& (SLO)
\end{tabular}
\caption{
Examples of GA (top) and RPD (bottom) segmentations from different models using the 5\% and the 20\% of the training data, respectively.
\textcolor{OliveGreen}{True positives} are depicted in green; true negatives, in black; \textcolor{BrickRed}{false positives}, in red; and \textcolor{Blue}{false negatives}, in blue.
}%
\label{fig:seg_examples}
\end{figure}

\paragraph{Baseline comparison.}
We compared our approach to current state-of-the-art methods (IPN~\cite{Li_IPN_TMI_2020}, IPNv2~\cite{Li_IPNv2_arXiv_2020}, Lachinov \textit{et al}.~\cite{Lachinov_MICCAI_2021}, and ReSensNet~\cite{Seebok_OR_2022}), showing that we greatly improve the state of the art in GA segmentation in scenarios with limited labeled data (Fig.~\ref{fig:results_models}).
When using only 5\% of the data (40 samples), the mean Dice score was 23\% higher than the best state-of-the-art approach.
Even without SSL, the proposed CNN improves the Dice score by 8\%.
This gain is even greater in terms of Area diff.
The improvement is also visible in the predicted segmentation masks (Fig.~\ref{fig:seg_examples}).
When using our approach, the number of false positives and negatives is highly reduced.
On the other hand, the improvement for RPD segmentation is more modest (in this case, we only compared with the current state-of-the-art-method: Lachinov et al.~\cite{Lachinov_MICCAI_2021}).
This can be explained by the greater visibility of GA features compared to RPD features in FAF and SLO (see Figs. \ref{fig:clinical_background} and \ref{fig:seg_examples}).
This suggests that SSL benefits from images with similar pathomorphological manifestations.

\paragraph{SSL effect.} To further assess the effect of the SSL, we also applied the strategy to the CNN by Lachinov et. al~\cite{Lachinov_MICCAI_2021}. Fig.~\ref{fig:results_models} shows that SSL clearly improves the GA and RPD segmentation performance of both proposed and Lachinov \textit{et al}. methods.
These results are in line with the qualitative results in Fig.~\ref{fig:seg_examples}.
This demonstrates that the SSL strategy is beneficial regardless of the architecture and the data.
Notwithstanding, as discussed in the baseline comparison, the proposed SSL is more beneficial for GA segmentation than for RPD.

\paragraph{Reconstructed modality effect.}
We also conducted experiments to assess the effect of the reconstructed modality (SLO and FAF).
Fig.~\ref{fig:results_models} shows that using FAF for SSL usually leads to better segmentation performance than using SLO.
However, in multiple cases, the differences were not statistically significant.
This is important because, unlike FAF, SLO does not require an external registration method.

\section{Conclusions}

Labeled data scarcity is one of the main limiting factors for the application of deep learning in medical imaging.
In this work, we have proposed a new model and SSL strategy for label-efficient 3D$\rightarrow$2D segmentation.
The proposed approach was validated in two tasks with clinical relevance: the en-face segmentation of GA and RPD in OCT.
The results demonstrate that:
\begin{inparaenum}[(1)]
\item the proposed CNN architecture clearly outperforms the state of the art when there is limited annotated data,
\item regardless of the architecture and the modality to be reconstructed, the proposed SSL strategy improves the performance of the models on the target tasks in those cases;
\item despite the greater diagnostic utility of FAF over SLO, SSL with FAF does not always result in a significant gain in model performance, with the advantage of the latter not requiring a supplementary registration method.
\end{inparaenum}
On the other hand, although the proposed approach shows promising results in the en-face segmentation of RPD, further evaluation is needed.

Based on our findings, we believe that the proposed approach has the potential to be used in other common 3D\ra2D tasks, such as the prediction of retinal sensitivity in OCT, the segmentation of different structures in OCT-A, or the segmentation of intravascular ultrasound (IVUS).
In addition, we also believe that the proposed SSL strategy could be easily extended to other imaging domains, such as magnetic resonance, where multi-modal data is widely used.


\subsubsection{Acknowledgements.}%
This work was supported in part by the Christian Doppler Research Association, Austrian Federal Ministry for Digital and Economic Affairs, the National Foundation for Research, Technology and Development, and Heidelberg Engineering.

%
%


\bibliographystyle{splncs04}

\bibliography{bibliography}

\clearpage


\appendix

\section*{Supplementary material}


\setlength{\tabcolsep}{8pt}

\begin{table}[tbph]
\centering
\caption{%
Hardware and software of the machine used to train and evaluate the models.
}%
\label{tab:hw_sw}
\begin{tabular}{ll}
\toprule
\multicolumn{2}{c}{\textbf{Hardware}} \\
\midrule
\textbf{Component} &  \textbf{Details} \\
\midrule

GPU &  NVIDIA RTX A6000, 48GB \\

CPU &  AMD EPYC 7302 16-Core Processor \\

Memory &  DDR4, 512GB \\

\bottomrule

\end{tabular}

\vspace{0.7cm}

\begin{tabular}{ll}
\toprule
\multicolumn{2}{c}{\textbf{Software}} \\
\midrule
\textbf{Type} &  \textbf{Details} \\
\midrule

Operating system &  CentOS Linux 7 (Core) \\

\midrule

\multirow{2}{*}{GPU} & CUDA V11.7.64 \\

& NVIDIA Driver 515.43.04 \\

\midrule

\multirow{4}{*}{Programming language and libraries} & Python 3.6.8 \\

& PyTorch 1.10.2+cu113 \\

& Torchvision 0.11.3+cu113 \\

& PyTorch Lightning 1.5.10 \\

\bottomrule

\end{tabular}

\end{table}


\begin{table}[tbph]
\centering
\caption{%
Approximate times and memory footprints of the proposed approach, measured on a machine with the hardware and software shown in Table~\ref{tab:hw_sw}.
For geographic atrophy (GA) and reticular pseudodrusen (RPD) segmentation, the measurements were taken using 100\% of the training data.
For all the other cases, the measurements are the same, except for the training time, which is approximately proportional to the training data.
}%
\label{tab1}
\resizebox{\textwidth}{!}{%
\begin{tabular}{lcccc}
\toprule
\textbf{Stage} &  \textbf{Training time} & \textbf{Training memory} & \textbf{Inference time} & \textbf{Inference memory} \\
 &  & \textbf{footprint} & \textbf{(per sample)} & \textbf{footprint} \\
\midrule

Reconstruction & 23h & 12GB & $<$1s & 5GB \\

GA segmentation & 3h & 24GB & $<$1s & 5GB \\

RPD segmentation & 0.6h & 24GB & $<$1s & 5GB \\
\bottomrule

\end{tabular}
}
\end{table}


\setlength{\tabcolsep}{4pt}

\begin{table}[tbph]
\centering
    \caption{
        GA segmentation results of the different methods in the different datasets.
        The best automatic results are highlighted in bold.
        A Wilcoxon signed-rank test of our Proposed (FAF) model vs. the others was performed in each case.
        (*: $p < 0.05$, **: $p < 0.01$, ***: $p < 0.001$).
    }

\resizebox{\textwidth}{!}{%
\begin{tabular}{lllllllll}

\toprule
\multicolumn{9}{c}{\textbf{GA-S-2}} \\
\midrule
  Method $\downarrow$ / Ratio $\rightarrow$ & \multicolumn{2}{c}{0.05} & \multicolumn{2}{c}{0.1} & \multicolumn{2}{c}{0.2} & \multicolumn{2}{c}{1.0} \\
  \cmidrule(lr){2-3}\cmidrule(lr){4-5}\cmidrule(lr){6-7}\cmidrule(lr){8-9}
  & Dice & Area diff. & Dice & Area diff. & Dice & Area diff. & Dice & Area diff. \\
\midrule

  IPN
       & $67.60 \pm 21.83$\textsuperscript{***} & $4.40 \pm 2.85$\textsuperscript{***}
       & $76.00 \pm 18.20$\textsuperscript{***} & $2.09 \pm 1.95$\textsuperscript{*}
       & $80.28 \pm 15.62$\textsuperscript{***} & $1.58 \pm 1.91$
       & $82.67 \pm 15.73$\textsuperscript{*} & $0.80 \pm 1.06$
       \\
  \rowcolor{lg}
  IPNv2
       & $71.33 \pm 18.93$\textsuperscript{***} & $2.29 \pm 1.91$\textsuperscript{*}
       & $81.99 \pm 12.57$\textsuperscript{***} & $1.16 \pm 1.00$
       & $84.19 \pm 11.32$\textsuperscript{*} & $0.84 \pm 0.95$
       & $85.71 \pm 11.69$\textsuperscript{*} & $0.91 \pm 1.19$
       \\
  Lachinov et al.
       & $68.94 \pm 23.59$\textsuperscript{***} & $4.47 \pm 3.44$\textsuperscript{***}
       & $81.95 \pm 15.32$\textsuperscript{*} & $1.43 \pm 1.40$
       & $83.96 \pm 12.99$\textsuperscript{*} & $1.18 \pm 0.97$\textsuperscript{*}
       & $87.77 \pm 9.41$ & $0.78 \pm 0.71$
       \\
  \rowcolor{lg}
  Lachinov et al. (SLO)
       & $73.36 \pm 20.72$\textsuperscript{***} & $3.38 \pm 2.37$\textsuperscript{***}
       & $80.47 \pm 16.85$\textsuperscript{*} & $1.77 \pm 1.56$\textsuperscript{*}
       & $83.04 \pm 14.06$\textsuperscript{*} & $1.50 \pm 1.15$\textsuperscript{*}
       & $86.93 \pm 11.04$ & $1.12 \pm 0.92$\textsuperscript{*}
       \\
  ReSensNet
       & $77.55 \pm 15.74$\textsuperscript{***} & $0.91 \pm 0.64$
       & $83.27 \pm 11.15$\textsuperscript{*} & $0.86 \pm 0.78$
       & $86.05 \pm 10.39$ & $0.75 \pm 0.72$
       & $\mathbf{88.27 \pm 8.12}$ & $\mathbf{0.71 \pm 0.50}$
       \\
  \rowcolor{lg}
  Proposed
       & $79.69 \pm 15.72$\textsuperscript{***} & $\mathbf{0.79 \pm 0.73}$
       & $83.87 \pm 11.22$\textsuperscript{*} & $\mathbf{0.68 \pm 0.52}$
       & $87.10 \pm 8.71$ & $\mathbf{0.68 \pm 0.40}$
       & $87.85 \pm 8.88$ & $0.92 \pm 0.58$\textsuperscript{*}
       \\
  Proposed (SLO)
       & $85.32 \pm 9.20$\textsuperscript{*} & $1.05 \pm 0.78$\textsuperscript{*}
       & $\mathbf{87.99 \pm 8.19}$ & $0.74 \pm 0.59$
       & $\mathbf{87.77 \pm 8.50}$ & $0.69 \pm 0.64$
       & $87.77 \pm 9.03$\textsuperscript{*} & $0.89 \pm 0.61$\textsuperscript{*}
       \\
  \rowcolor{lg}
  Proposed (FAF)
       & $\mathbf{86.26 \pm 9.36}$ & $0.85 \pm 0.73$
       & $87.00 \pm 9.41$ & $0.91 \pm 0.71$
       & $87.30 \pm 8.84$ & $0.85 \pm 0.66$
       & $88.10 \pm 8.68$ & $0.82 \pm 0.59$
       \\
  \textit{2nd observer}
       & $91.56 \pm 7.22$ & $0.61 \pm 0.51$
       & $91.56 \pm 7.22$ & $0.61 \pm 0.51$
       & $91.56 \pm 7.22$ & $0.61 \pm 0.51$
       & $91.56 \pm 7.22$ & $0.61 \pm 0.51$
       \\

       \bottomrule

       &&&&& \\
       &&&&& \\
       &&&&& \\

       \toprule
\multicolumn{9}{c}{\textbf{GA-S-1}} \\
\midrule
  Method $\downarrow$ / Ratio $\rightarrow$ & \multicolumn{2}{c}{0.05} & \multicolumn{2}{c}{0.1} & \multicolumn{2}{c}{0.2} & \multicolumn{2}{c}{1.0} \\
  \cmidrule(lr){2-3}\cmidrule(lr){4-5}\cmidrule(lr){6-7}\cmidrule(lr){8-9}
  & Dice & Area diff. & Dice & Area diff. & Dice & Area diff. & Dice & Area diff. \\
       \midrule

  IPN
       & $71.45 \pm 19.72$\textsuperscript{***} & $3.84 \pm 2.94$\textsuperscript{***}
       & $82.44 \pm 14.77$\textsuperscript{***} & $1.34 \pm 1.38$\textsuperscript{***}
       & $86.03 \pm 11.60$\textsuperscript{***} & $0.89 \pm 0.75$\textsuperscript{***}
       & $90.26 \pm 7.48$\textsuperscript{***} & $0.54 \pm 0.49$\textsuperscript{***}
       \\
  \rowcolor{lg}
  IPNv2
       & $74.22 \pm 17.46$\textsuperscript{***} & $2.88 \pm 3.65$\textsuperscript{***}
       & $84.90 \pm 12.94$\textsuperscript{***} & $0.93 \pm 0.84$\textsuperscript{***}
       & $87.63 \pm 9.78$\textsuperscript{***} & $0.71 \pm 0.58$\textsuperscript{***}
       & $91.57 \pm 6.62$\textsuperscript{***} & $0.51 \pm 0.55$\textsuperscript{**}
       \\
  Lachinov et al.
       & $73.46 \pm 21.50$\textsuperscript{***} & $4.10 \pm 3.50$\textsuperscript{***}
       & $86.89 \pm 13.75$\textsuperscript{***} & $1.02 \pm 1.44$\textsuperscript{***}
       & $90.20 \pm 9.98$\textsuperscript{***} & $0.62 \pm 0.68$\textsuperscript{**}
       & $92.89 \pm 5.48$\textsuperscript{**} & $0.38 \pm 0.40$
       \\
  \rowcolor{lg}
  Lachinov et al. (SLO)
       & $81.78 \pm 17.68$\textsuperscript{***} & $2.09 \pm 2.13$\textsuperscript{***}
       & $88.45 \pm 13.78$\textsuperscript{***} & $0.91 \pm 1.25$\textsuperscript{**}
       & $90.98 \pm 9.58$\textsuperscript{**} & $0.62 \pm 0.55$\textsuperscript{***}
       & $92.82 \pm 7.31$ & $0.45 \pm 0.54$\textsuperscript{*}
       \\
  ReSensNet
       & $78.47 \pm 17.44$\textsuperscript{***} & $1.55 \pm 1.52$\textsuperscript{***}
       & $83.81 \pm 16.02$\textsuperscript{***} & $1.06 \pm 1.04$\textsuperscript{***}
       & $89.94 \pm 7.31$\textsuperscript{***} & $0.62 \pm 0.51$\textsuperscript{***}
       & $92.70 \pm 7.22$\textsuperscript{*} & $\mathbf{0.34 \pm 0.30}$
       \\
  \rowcolor{lg}
  Proposed
       & $79.90 \pm 15.51$\textsuperscript{***} & $1.62 \pm 1.69$\textsuperscript{***}
       & $84.51 \pm 17.70$\textsuperscript{***} & $0.97 \pm 0.80$\textsuperscript{***}
       & $90.90 \pm 7.76$\textsuperscript{***} & $0.56 \pm 0.57$\textsuperscript{*}
       & $93.15 \pm 5.67$ & $0.44 \pm 0.37$\textsuperscript{***}
       \\
  Proposed (SLO)
       & $88.36 \pm 8.98$\textsuperscript{***} & $0.85 \pm 1.25$\textsuperscript{***}
       & $91.40 \pm 7.49$ & $\mathbf{0.45 \pm 0.44}$
       & $91.48 \pm 6.86$\textsuperscript{***} & $0.47 \pm 0.80$
       & $92.93 \pm 5.72$\textsuperscript{***} & $0.39 \pm 0.35$\textsuperscript{*}
       \\
  \rowcolor{lg}
  Proposed (FAF)
       & $\mathbf{90.47 \pm 8.33}$ & $\mathbf{0.53 \pm 0.53}$
       & $\mathbf{91.61 \pm 6.52}$ & $0.48 \pm 0.38$
       & $\mathbf{92.21 \pm 6.08}$ & $\mathbf{0.42 \pm 0.37}$
       & $\mathbf{93.15 \pm 5.71}$ & $0.36 \pm 0.32$
       \\

       \bottomrule

       &&&&& \\
       &&&&& \\
       &&&&& \\

\toprule
\multicolumn{9}{c}{\textbf{GA-M-S}} \\
\midrule
  Method $\downarrow$ / Ratio $\rightarrow$ & \multicolumn{2}{c}{0.05} & \multicolumn{2}{c}{0.1} & \multicolumn{2}{c}{0.2} & \multicolumn{2}{c}{1.0} \\
  \cmidrule(lr){2-3}\cmidrule(lr){4-5}\cmidrule(lr){6-7}\cmidrule(lr){8-9}
  & Dice & Area diff. & Dice & Area diff. & Dice & Area diff. & Dice & Area diff. \\
\midrule

  IPN
       & $62.29 \pm 29.57$\textsuperscript{***} & $5.55 \pm 5.38$\textsuperscript{***}
       & $70.38 \pm 28.41$\textsuperscript{***} & $3.02 \pm 4.09$\textsuperscript{***}
       & $74.59 \pm 25.77$\textsuperscript{***} & $1.77 \pm 1.78$\textsuperscript{***}
       & $80.32 \pm 23.25$\textsuperscript{***} & $1.01 \pm 1.92$\textsuperscript{*}
       \\
  \rowcolor{lg}
  IPNv2
       & $61.31 \pm 27.32$\textsuperscript{***} & $5.66 \pm 5.11$\textsuperscript{***}
       & $70.94 \pm 25.45$\textsuperscript{***} & $2.35 \pm 2.44$\textsuperscript{***}
       & $72.87 \pm 24.43$\textsuperscript{***} & $2.38 \pm 2.81$\textsuperscript{***}
       & $76.30 \pm 26.29$\textsuperscript{***} & $1.48 \pm 1.57$\textsuperscript{***}
       \\
  Lachinov et al.
       & $62.50 \pm 30.87$\textsuperscript{***} & $6.20 \pm 5.72$\textsuperscript{***}
       & $71.92 \pm 30.08$\textsuperscript{***} & $3.09 \pm 3.90$\textsuperscript{***}
       & $78.57 \pm 27.69$\textsuperscript{***} & $1.21 \pm 1.84$\textsuperscript{**}
       & $\mathbf{84.89 \pm 19.96}$ & $0.43 \pm 0.33$
       \\
  \rowcolor{lg}
  Lachinov et al. (SLO)
       & $71.60 \pm 31.44$\textsuperscript{***} & $3.29 \pm 4.32$\textsuperscript{***}
       & $75.71 \pm 29.89$\textsuperscript{**} & $2.16 \pm 3.21$\textsuperscript{*}
       & $81.15 \pm 24.67$\textsuperscript{***} & $0.86 \pm 1.34$\textsuperscript{**}
       & $84.20 \pm 21.81$ & $0.42 \pm 0.49$
       \\
  ReSensNet
       & $72.31 \pm 23.16$\textsuperscript{***} & $1.99 \pm 2.21$\textsuperscript{***}
       & $73.30 \pm 24.82$\textsuperscript{***} & $1.47 \pm 1.97$\textsuperscript{**}
       & $81.49 \pm 22.24$\textsuperscript{***} & $0.82 \pm 1.23$\textsuperscript{**}
       & $80.06 \pm 23.02$\textsuperscript{***} & $0.90 \pm 1.16$\textsuperscript{***}
       \\
  \rowcolor{lg}
  Proposed
       & $73.03 \pm 21.17$\textsuperscript{***} & $1.62 \pm 2.02$\textsuperscript{***}
       & $76.53 \pm 23.78$\textsuperscript{***} & $1.16 \pm 1.34$\textsuperscript{**}
       & $83.33 \pm 21.30$\textsuperscript{***} & $0.57 \pm 0.54$\textsuperscript{*}
       & $84.85 \pm 20.88$ & $\mathbf{0.38 \pm 0.29}$
       \\
  Proposed (SLO)
       & $82.47 \pm 21.11$\textsuperscript{***} & $0.54 \pm 0.57$\textsuperscript{*}
       & $82.57 \pm 19.64$ & $0.69 \pm 0.83$
       & $83.36 \pm 20.34$\textsuperscript{***} & $0.56 \pm 0.54$\textsuperscript{**}
       & $83.67 \pm 22.15$\textsuperscript{*} & $0.41 \pm 0.37$
       \\
  \rowcolor{lg}
  Proposed (FAF)
       & $\mathbf{84.59 \pm 19.66}$ & $\mathbf{0.37 \pm 0.31}$
       & $\mathbf{82.80 \pm 21.18}$ & $\mathbf{0.52 \pm 0.44}$
       & $\mathbf{85.35 \pm 19.23}$ & $\mathbf{0.35 \pm 0.35}$
       & $84.34 \pm 21.33$ & $0.41 \pm 0.35$
       \\       
       \bottomrule

\end{tabular}%
}
\end{table}


\begin{table}[tbph]
\centering
    \caption{
        RDP segmentation results of the different methods in the RPD-S dataset.
        The best results from a model are highlighted in bold.
        A Wilcoxon signed-rank test of our Proposed (SLO) model vs. the others was performed in each case, but no statistically significant differences were found.
    }

\begin{tabular}{lllll}

\toprule
\multicolumn{5}{c}{\textbf{RPD-S}} \\
\midrule
  Method $\downarrow$ / Ratio $\rightarrow$ & \multicolumn{2}{c}{0.2} & \multicolumn{2}{c}{1.0} \\
  \cmidrule(lr){2-3}\cmidrule(lr){4-5}
  & Dice & Area diff. & Dice & Area diff. \\
  \midrule

  Lachinov et al.
       & $41.34 \pm 7.98$ & $2.42 \pm 1.75$
       & $\mathbf{51.62 \pm 11.07}$ & $2.40 \pm 1.99$
       \\
  \rowcolor{lg}
  Lachinov et al. (SLO)
       & $42.24 \pm 8.84$ & $\mathbf{2.32 \pm 1.66}$
       & $51.45 \pm 9.96$ & $1.97 \pm 1.37$
       \\
  Proposed
       & $45.47 \pm 10.31$ & $2.56 \pm 2.12$
       & $50.60 \pm 12.57$ & $3.46 \pm 2.66$
       \\
  \rowcolor{lg}
  Proposed (SLO)
       & $\mathbf{47.55 \pm 12.65}$ & $2.84 \pm 3.49$
       & $48.28 \pm 17.24$ & $\mathbf{1.94 \pm 1.37}$
       \\
       
\bottomrule

\end{tabular}%

\end{table}

\end{document}